\documentclass{article}
\usepackage{graphicx}

\begin{document}

\title{Scattering of Solitons and Dark Solitons by Potential Walls in the Nonlinear Schr\"odinger Equation}
\author{Hidetsugu Sakaguchi and Mitsuaki Tamura\\
\\
Department of Applied Science for Electronics and Materials,\\ Interdisciplinary Graduate School of Engineering Sciences,\\
 Kyushu University, Kasuga, Fukuoka 816-8580, Japan}
\maketitle

\begin{abstract}
Scattering of solitons and dark solitons by potential walls is studied in the nonlinear Schr\"odinger equation under various conditions. We investigate the conditions under which solitons are split into two solitons at the potential wall. We find that a soliton can be trapped in an interspace between two potential walls. A dark soliton can also be scattered at the potential wall.  Similarly to a bright soliton, a dark soliton can pass through more easily the potential wall, as the width of the dark soliton is larger. A dark soliton can run away spontaneously from an interspace between the two potential walls. We also study the motion of a two-dimensional soliton in a two-dimensional quintic nonlinear Schr\"odinger equation. We find the coherent tunneling through a potential wall, and the refraction corresponding to Newton's refraction theory.
\end{abstract}
\section{Introduction}
The nonlinear Schr\"odinger equation is a typical soliton equation.\cite{Karp} Solitons in optical fibers are described by the nonlinear Schr\"odinger equation and have been intensively studied for the optical communication.\cite{Hase,Kiv} Recently, solitons and dark solitons were observed in the Bose-Einstein condensates (BECs).\cite{Denschlag, Burger,Strecker,Khay} Dark (bright) solitons are found in the BECs with repulsive (attractive) interaction.  
The nonlinear Schr\"odinger equation is called the Gross-Pitaevskii equation in the research field of BECs. In the BECs, matter waves obey the nonlinear Schr\"odinger equation with external potentials. The harmonic potentials are used to confine the BECs. Interference patterns of laser beams generate external periodic potentials. Solitons in external potentials have been studied by several authors.\cite{Moura,Frau,Trom} Especially, solitons in periodic potentials (optical lattices) are intensively studied.\cite{Baiz,Sakaguchi}
 However, the scattering of a soliton by a potential wall was not studied in detail, although it is a simple and a fundamental process.  The scattering problem of a quantum particle by the potential wall, which is described by the linear Schr\"odinger equation, is a typical problem of the quantum mechanics. In our previous letter, we studied the scattering and trapping of solitons by a potential wall and well. We found that the soliton behaves like a classical mechanical particle under a certain condition, and evaluate the critical energies for the transmission. In this paper, we study a transition between the classical mechanical behavior and the wavy behavior more in detail. We will study also the scattering of a dark soliton by the potential wall. We will further study the scattering of a two-dimensional soliton. 
\section{Scattering of bright solitons by a potential wall}

The nonlinear Schr\"odinger equation with a potential term is written as 
\begin{equation}
i\frac{\partial \phi}{\partial t}=-\frac{1}{2}\frac{\partial^2\phi}{\partial x^2}+g|\phi|^2\phi+U(x)\phi,
\end{equation}
where $\phi(x,t)$ is the wave function, $g$ is the coefficient of the nonlinear term, $U(x)$ is an external potential. First, we consider the case of $g=-1$.  It corresponds to the case of the attractive interaction in the problem of BECs. When $U(x)=0$, the nonlinear Schr\"odinger equation with $g=-1$ has a soliton solution
\begin{equation}
\phi(x,t)=A{\rm sech}\{A(x-vt)\}e^{ip(x-vt)-i\omega t},
\end{equation}
where $v=p$ and $\omega=-A^2/2+v^2/2$. 

In this section, we study the scattering by a simplest potential wall, which is expressed as
\begin{equation}
U(x)=U_0,\;\;{\rm for}\;\;x_0<x<x_0+d,\;\;\;
U(x)=0,\;\;{\rm for}\;\; x<x_0, \;x>x_0+d,
\end{equation}
where $d$ is the width of the potential wall and $U_0>0$ is the height. 
We study numerically the scattering of a soliton by the potential wall.  We perform numerical simulations using the split-step Fourier method with Fourier modes 1024 or 2048. The periodic boundary conditions are assumed, and the time step is 0.001. Figure 1(a) displays the time evolution of a soliton passing through a potential wall with $U_0=0.7$ and $d=1$. The system size is $L=100$ and the potential wall locates at $x=x_0=L/2=50$. The initial velocity of the soliton is $v=1$ and the amplitude is $A=1.5$. The soliton is split into two pulses by the potential wall. The transmission coefficient is 0.488, which is calculated as $T=\int_{L/2}^L|\phi|^2dx/\int_0^L|\phi|^2dx$ after the scattering. 
Figure 1(b) displays the transmission coefficient $T$ as a function of the initial velocity $v$ for three values of $U_0$ for $A=1.5$ and $d=1$.
For $U_0=0.1$, the transmission coefficient changes from 0 to 1 abruptly. 
The soliton is completely reflected below the critical value of $v$, and completely transmitted above the critical velocity. The soliton behaves like a classical mechanical particle. We showed that the critical kinetic energy depends strongly on $A$ as $E_c=U_0\tanh(Ad/2)$ using the Lagrangian formalism.\cite{Sakaguchi} For $U=0.7$, the transmission coefficient changes smoothly from 0 to 1. 
This behavior is characteristic of the transmission of waves or the quantum mechanical particle. 
Near $U=0.4$, the scattering changes from the particlelike behavior to the wavy behavior for the amplitude $A=1.5$.  The transmission coefficient changes in a non-monotonic manner near the critical velocity, although we do not understand the reason yet. 
We discussed qualitatively the transition using the conservation law of the energy in the previous paper. We confirm the conjecture with the direct numerical simulations. 
\begin{figure}[t]
\begin{center}
\includegraphics[width=13cm]{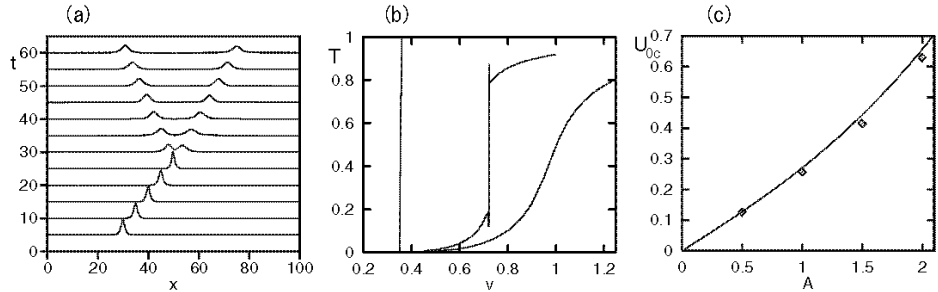}
\end{center}
\caption{(a) Time evolution of $|\phi(x,t)|$ for $U_0=0.7,A=1.5$ and $v=1$.
A soliton is splitted into two pulses by the potential wall of $d=1$ located at $x=L/2=50$. (b) Transmission coefficient $T$ as a function of the initial velocity $v$. (c) A boundary line between the particlelike behavior and the wavy behavior. Under the line, a soliton behaves like a particle. The solid line is the theoretical estimate and the rhombi denote the critical value of $U_0$ obtained numerically.}
\label{f1}
\end{figure}
\begin{figure}[t]
\begin{center}
\includegraphics[width=13cm]{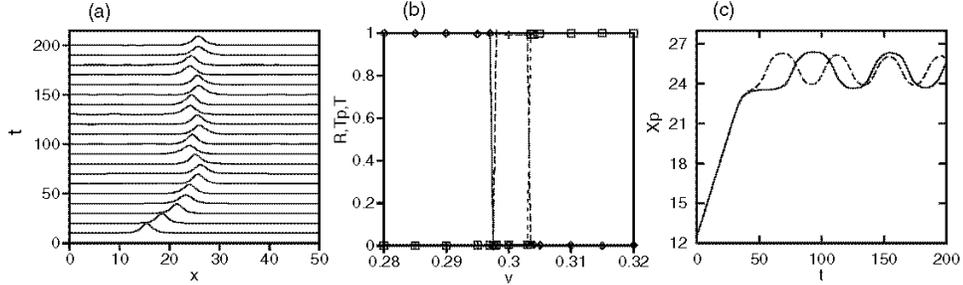}
\end{center}
\caption{(a) Time evolution of a soliton for $U_0=0.1,\,U_1=0,\,A=1$ and $v=0.302$.
A soliton penetrates into a potential wall, and then is trapped between the double walls. The walls located at $x_1=23$ and $x_2=26$. (b) Reflection coefficient $R$ (solid line with rhombi), the trapping coefficient $T_p$ (dashed line with crosses) and the transmission coefficient $T$ (dotted line with squares) as a function of the initial velocity $v$. (c) Time evolution of the peak position of the soliton (dashed line) and the theoretical model of eq.~(6) (solid line).}
\label{f1}
\end{figure}
\begin{figure}[t]
\begin{center}
\includegraphics[width=15cm]{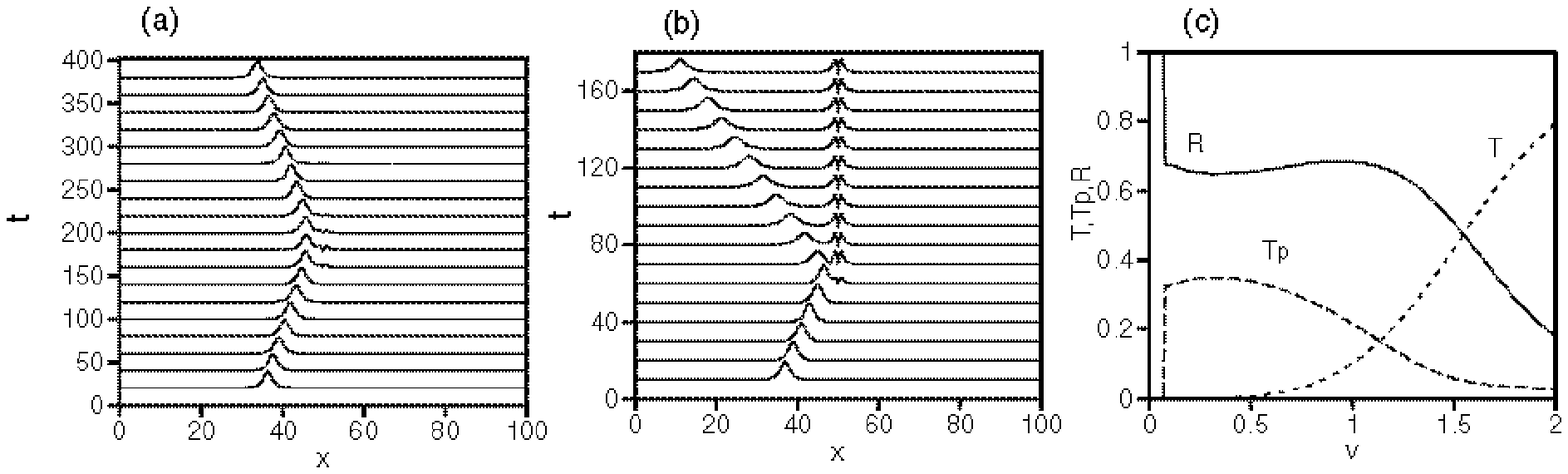}
\end{center}
\caption{(a) Reflection process of a soliton for $U_0=0.6,\,U_1=2.5,\,A=1$ and $v=0.07$.
(b) Generation of a bound state at the collision for $U_0=0.6,\,U_1=2.5,\,A=1$ and $v=0.2$. (c) Reflectivity $R$ (solid line), trapping rate $T_p$ (dashed line) and transmissivity $T$ (dotted line) as a function of $v$ for $U_0=0.6,\,U_1=2.5$ and $A=1$.}
\label{f3}
\end{figure}

We first explain a criterion based on the conservation laws.
We assume that a soliton is located initially at $x=x_0+d/2$ with velocity 0.  
If the height $U_0$ of the potential wall is sufficiently large, the soliton will be split into two pulses.  However, the soliton cannot be split, if $U_0$ is not so large owing to the attractive mutual interaction.  Equation (1) has at least two invariants, that is, the total mass $N=\int_0^L|\phi|^2dx$ and the total energy $E=\int_0^L\{1/2|\phi_x|^2-1/2|\phi|^4+U(x)|\phi|^2\}dx$.  If the soliton is split into two counterpropagating solitons: $\phi(x,t)=A^{\prime}{\rm sech}\{A^{\prime}(x-vt-L/2)\}e^{ip(x-vt-L/2)-i\omega t}+A^{\prime}{\rm sech}\{A^{\prime}(x+vt-L/2)\}e^{-ip(x+vt-L/2)-i\omega t}$, the amplitude $A^{\prime}$ and the momentum $p$ satisfy $A^{\prime}=A/2$ and $2p^2A^{\prime}=-1/4A^3+2U_0A\tanh(Ad/2)$
 owing to the conservation laws. If $U_0<A^2/\{8\tanh(Ad/2)\}$, there is no real solution for $p$, that is, the splitting cannot occur and the one-pulse structure is maintained.  Above the line $U_0=A^2/\{8\tanh(Ad/2)\}$, the potential energy is used to split a soliton against the attractive interaction. That is, the above argument suggests that a criterion between keeping one-pulse state and splitting into two pulses is $U_{0c}=A^2/\{8\tanh(Ad/2)\}$.  To check the conjecture of the criterion, we have performed numerical simulations, changing the value of $U_0$ for several $A$'s, and constructed the same type plots as Fig.~1(b). We have regarded the critical value of $U_c$ for each $A$ as the lowest value of $U_0$, at which the transmission rate $T$ has a jump across $T=0.5$.  Figure 1(c) displays the critical values $U_{0c}$ for $A=0.5,1,1.5$ and 2. The dashed line is the theoretical estimate $U_{0c}=A^2/\{8\tanh(Ad/2)\}$. The transition between the particlelike behavior to the wavy behavior is not so clear-cut, and so the above theoretical curve gives a fairly good criterion. 

The splitting of the one-soliton was caused by the perturbation at the potential wall. It is desirable to determine the reflection and transmission coefficients theoretically using the inverse scattering method or other perturbative methods but we have not succeeded in it yet. 

\section{Scattering of bright solitons by double potential walls}
In this section, we consider the scattering of bright solitons  by a potential with double walls:
\begin{eqnarray}
U(x)&=&U_0,\;\;{\rm for}\;\;x_1<x<x_1+d,\nonumber\\
&=&-U_1,\;\;{\rm for}\;\;x_1+d<x<x_2,\nonumber\\
&=&U_0,\;\;{\rm for }\;\; x_2<x<x_2+d,\nonumber\\
&=&0,\;\;{\rm for }\;\; x<x_1,\;\;x>x_2+d.
\end{eqnarray}
There are two potential wall at $x=x_1$ and $x_2$. 
Figure 2(a) displays a time evolution of a soliton with $A=1$ and the initial velocity $v=0.302$. The system size is 50, and the parameters of the potential are $U_0=0.1$, $U_1=0$, $x_1=23$, $x_2=26$ and $d=1$. The soliton passes over the first potential wall, but is reflected by the second wall, and then trapped in an interspace between the two walls. Figure 2(b) displays the reflectivity $R$, the transmissivity $T$, and the trapping rate $T_p$, which are defined as $R=\int_{0}^{L/2-5}|\phi|^2dx/\int_0^L|\phi|^2dx$, $T_p=\int_{L/2-5}^{L/2+5}|\phi|^2dx/\int_0^L|\phi|^2dx$ and $T=\int_{L/2+5}^{L}|\phi|^2dx/\int_0^L|\phi|^2dx$ after the scattering. The reflectivity and the transmissivity changes from 0 to 1 abruptly, and the soliton behaves like a particle. The soliton is trapped between the double walls in a narrow parameter region of $0.298\le v\le 0.303$. This trapping phenomenon can be described also using the Lagrangian formalism.

The nonlinear Schr\"odinger equation is rewritten with the Lagrangian form: $\partial/\partial t(\delta L/\delta \phi_t)=\delta L/\delta \phi$,
where $L=\int_{-\infty}^{\infty} dx \{i/2(\phi_t\phi^*-\phi_t^*\phi)-1/2|\phi_x|^2+1/2|\phi|^4-U(x)|\phi|^2\}$.
As is used in the previous paper, we assume a solution of the form  
\begin{equation}
\phi(x,t)=A(t){\rm sech}\{(x-\xi(t))/a(t)\}e^{ip(t)(x-\xi(t))+i\sigma(t)\log \cosh\{(x-\xi(t))/a(t)\}-i\omega t},\end{equation}
where $A,a,p,\xi$ and $\sigma$ change in time, but $N=2A^2a$ is conserved in time. The term  $i\sigma(t)\log \cosh(x-\xi(t))/a(t)$ is a kind of chirp term.  Owing to the inclusion of this term, the width of the soliton changes in time and breathing motion becomes possible.  Some energy is given to the breathing motion and the kinetic energy may be decreased. Substitution of eq.~(5) into the Lagrangian and its spatial integration yields an effective Lagrangian
\[L_{eff}=N(p\xi_t-\sigma_t(1-\log 2)+\sigma a_t/(2a)-1/(6a^2)-p^2/2-\sigma^2/(6a^2)+N/(6a)-U_{eff}),\]
where $U_{eff}=U_0/2[\tanh\{(x_1+d-q(t))/a\}-\tanh\{(x_1-q(t))/a\}+\tanh\{(x_2+d-q(t))/a\}-\tanh\{(x_2-q(t))/a\}]$ for $U_1=0$. 
The equations of motion for $p(t),\xi(t),a(t)$ and $\sigma(t)$ are 
\begin{eqnarray}
\frac{dq}{dt}&=&p,\;\;
\frac{dp}{dt}=-\frac{\partial U_{eff}}{\partial q},\nonumber\\
\frac{da}{dt}&=&\frac{2\sigma}{3a},\;\;
\frac{d\sigma}{dt}=2a\left(\frac{1+\sigma^2}{3a^3}-\frac{N}{6a^2}-\frac{\partial U_{eff}}{\partial a}\right ).
\end{eqnarray}
This equation is derived from a Hamiltonian $H= p^2/2+U_{eff}+1/(6a^2)+\sigma^2/(3a^2)+N/(6a)$ as $dq/dt=\partial H/\partial p,\;dp/dt=-\partial 
H/\partial q,\; da/dt=\partial H/\partial \sigma,\;d\sigma /dt=-\partial H/\partial a$.
The classical mechanical energy $E=p^2/2+U_{eff}$ is not conserved in this equation, and the width $a$ changes in time. If the classical mechanical energy $E$ decays in time, the soliton can be trapped between  the double walls. Figure 2(c) compares the time evolutions of $q$ by eq.~(6) with the peak position of the soliton by eq.~(1) for the initial velocity $v=p=0.302$. The trapping phenomenon may be qualitatively explained by eq.~(6).   

When $U_0$ and $U_1$ become larger, a soliton is split into two or three solitons. We show an example of the scattering of a soliton leaving a bound state behind. Figure 3 displays the soliton scattering with amplitude $A=1$ at $U_0=0.6$ and $U_1=2.5$. The system size is 100. Figure 3(a) displays the reflection process of the soliton for the initial velocity $v=0.07$. For this velocity, the reflectivity is almost 1. But for $v=0.08$, the reflectivity decreases abruptly to 0.675, and a bound state is created in the potential well.  Figure 3(b) displays the reflection process leaving a bound state behind for $v=0.2$. The bound state has a norm 0.688 and is almost stationary. The bound state is antisymmetric around the center, therefore, the amplitude $|\phi|$ is 0 at the center. The velocity of the reflected pulse is $v\sim-0.34$, and it is larger than the initial velocity. That is, the reflecting pulse is accelerated by leaving a bound state with low energy behind. Figure 3(c) displays the reflectivity $R$, the transmissivity $T$ and the trapping rate $T_p$ as a function of the initial velocity. There is a jump of the reflectivity at $v=0.07$. The bound state is created abruptly for $v\ge 0.08$. The transmissivity $T$ increases continuously from $v\sim 0.5$. Near $v=1.5$, a soliton is split into three pulses, i.e.,a reflecting pulse, a bound state and the transmitted pulse.

\begin{figure}[t]
\begin{center}
\includegraphics[width=15cm]{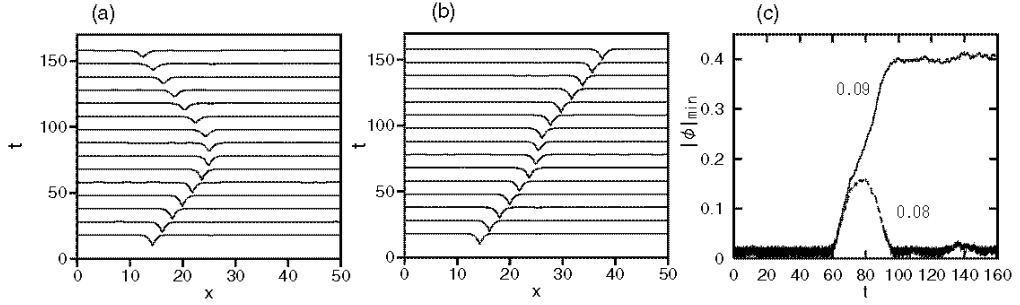}
\end{center}
\caption{(a) Time evolution of $|\phi(x,t)|$ for $U_0=0.09,A=1$ and $v=3\pi/50$. A dark soliton is reflected  by the potential wall of $d=1$ located at $x=L/2=25$. (b) Time evolution of $|\phi|$ for $U_0=0.08,A=1$ and $v=3\pi/50$.
A dark soliton passes over the potential wall. (c) Time evolution of the minimum value of $|\phi|$ for $U_0=0.09$ (solid line) and $U_0=0.08$ (dashed line).
}
\label{f4}
\end{figure}
\begin{figure}[t]
\begin{center}
\includegraphics[width=6cm]{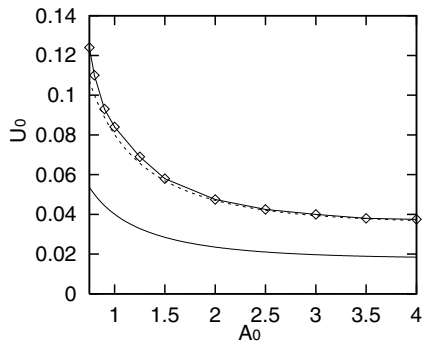}
\end{center}
\caption{The critical height $U_0$ of the potential wall, above which the dark soliton is reflected. The solid line is the theoretical estimate using the energy. The dashed line is the twice of that.}
\label{f5}
\end{figure}
\begin{figure}[t]
\begin{center}
\includegraphics[width=14cm]{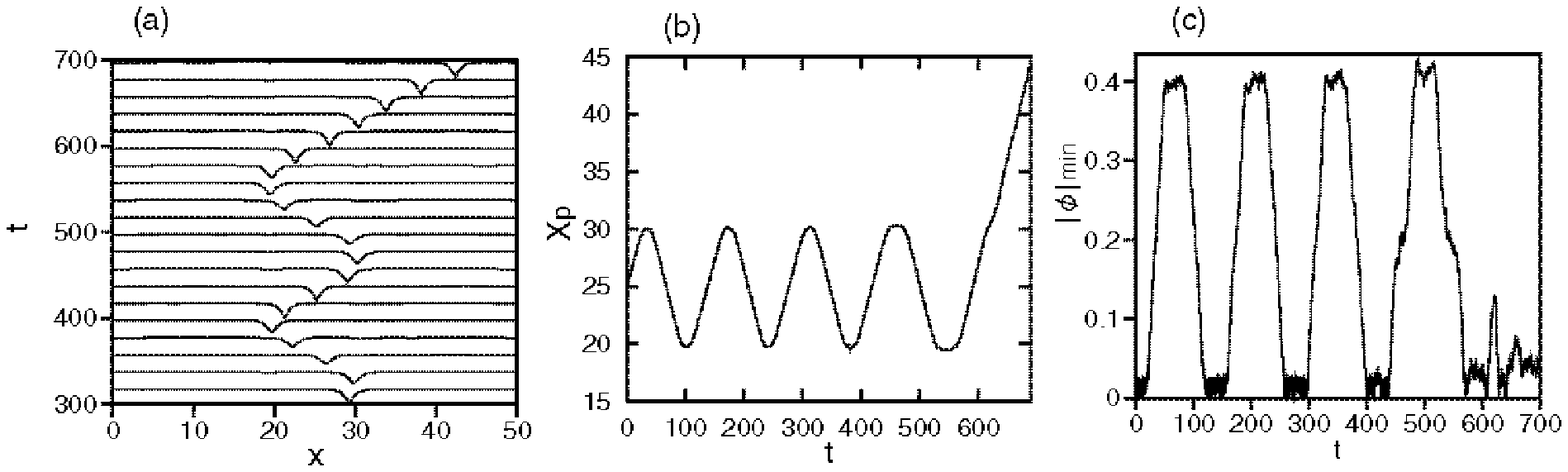}
\end{center}
\caption{(a) Time evolution of $|\phi(x,t)|$ for the double potential walls at $x_1=19$ and $x_2=30$ with $U_0=0.1,\,U_1=0,\,d=1$, $A=1$ and $v=3\pi/50$.
A dark soliton goes out from the interspace between the double walls after several reflections. 
 (b) Time evolution of the minimum point of $|\phi|$. (c) Time evolution of the minimum value of $|\phi|$.}
\label{f6}
\end{figure}
\section{Scattering of dark solitons}
If the interaction is repulsive in the BECs, the nonlinear Schr\"odinger equation has a form 
\begin{equation}
i\frac{\partial \phi}{\partial t}=-\frac{1}{2}\frac{\partial^2\phi}{\partial x^2}+|\phi|^2\phi+U(x)\phi,
\end{equation}
that is, the sign of the nonlinear term becomes positive and the coefficient is assumed to be 1. A dark soliton is a special solution to this nonlinear Schr\"odinger equation. A dark soliton has a form:
\begin{equation}
\phi(x,t)=\{iB+A\tanh A(x-vt)\}e^{ik(x-vt)+i(k^2/2+kB-A^2-B^2)t},
\end{equation}
where $B$, $A$ and $k$ are parameters which characterizes the dark soliton and the velocity $v$ is given by $k+B$. For $B=0$, the amplitude $|\phi|$ becomes 0 at the center. For $B\ne 0$, the amplitude does not become 0, and it is sometimes called a grey soliton. We consider the case of the potential given by eq.~(3). The energy of the dark soliton is given by $E=1/2\int dx\{|\partial \phi/\partial x|^2+|\phi|^4+2U(x)|\phi|^2\}=-2(A^2+B^2)A+4/3A^3-2kAB-k^2A-U_0A\{\tanh A(x_0+d-q)-\tanh A(x_0-q)\}$, where $q$ is the position of the dark soliton. 
We investigate numerically the time evolution of this dark soliton. The initial value of $B$ is set to be 0 and the initial velocity is given by $v=k$. The width of the potential wall in eq.~(3) is given by $d=1$, and the system size 50. In actual simulations, we have investigated the time evolution of two dark solitons (Another dark soliton is located at $q+L/2$) in a system of double size $L=100$, since we performs numerical simulations using the periodic boundary conditions. Owing the periodic boundary conditions, the initial wave number needs to be $k=2\pi n/L$ at $t=0$, where $n$ is an integer. Figure 4(a) displays the reflection process of the dark soliton by the potential wall for $U_0=0.09$, $v=k=3\pi/50$, and $A=1$. Figure 4(b) displays the transmission process for $U_0=0.08$, $v=3\pi/50$ and $A=1$. The dark soliton approaches plane waves in the infinity, and the scattering of the plane waves by the potential wall always occurs, before the center of the dark soliton reaches the potential wall. In Fig.~4(a) and (b), the amplitude $|\phi|$ is slightly depressed owing to the scattering  near the position $x=L/2$ of the potential wall, but the effect of the scattering of the plane waves seems to be small, when the potential height $U_0$ is small.    

Figure 4(c) displays the time evolution of the minimum value of $|\phi|$ for $U_0=0.09$ and 0.08. 
If the dark soliton keeps the form (8), the minimum value of $|\phi|$ represents $|B|$. The initial value of $B$ is assumed to be 0. In the case of the reflection, $B$ is decreased to -0.40, and the velocity of the reflected dark soliton is numerically estimated as -0.20, which is close to $v=k+B\sim -0.21$.  
The velocity of the reflected dark soliton is larger than the velocity $v=3\pi/50\sim 0.188$ of the initial dark soliton. Since the energy and the norm are conserved, $A^2+B^2=1$ and $-2A+4/3A^3-2kAB-k^2A=-2/3-k^2$. The $B$-value is expected as -0.404 from the two conservation laws if $k$ is fixed to be the initial value, and it is comparable to the numerical value. When the dark soliton passes over the potential wall, the $B$-value is decreased to -0.16 near the potential wall, but it recovers to 0 as the dark soliton goes away from the potential wall. 

There is a critical value of $U_0$ for the reflection to occur. Figure 5 displays the critical value of $U_0$ as a function of the initial value $A_0$ of $A$ above which the reflection is numerically observed for the initial velocity $v=3\pi/50$. The critical value increases as $A_0$ is decreased. It means that the dark soliton penetrates more easily the potential wall, as the amplitude is decreased. This effect is similar to the one we have already seen for bright solitons as coherent tunneling. The critical value may be similarly evaluated using the energy conservation law, if the form of the dark soliton is assumed to be given by eq.~(8). At the critical value of $U_0$, the dark soliton will stop at $x=x_0+d/2$. Then, the $B$-value is expected to be $-k$, since the velocity $v=k+B$ is 0. 
Owing to the energy conservation law, $-2/3A_0^3-k^2A_0=-2A_0^2A+4/3A^3+k^2A-2U_0A\tanh (Ad/2)$ is satisfied. The critical vale of $U_0$ is evaluated as $U_0=
(-2A_0^2A+4/3A^3+k^2A+k^2A_0+2/3A_0^3)/(2A\tanh Ad/2)$. 

A similar result can be derived from the Lagrangian formalism. 
We assume a solution of the form
\[
\phi(x,t)=\{iB(t)+A(t)\tanh A(t)(x-q(t))\}e^{ik(x-q(t))+i(k^2/2+kB-A^2-B^2)t}.\]
Substitute it to the Lagrangian
\[L=\int\{i/2(\phi_t\phi^*-\phi_t^*\phi)-1/2|\partial \phi/\partial x|^2-1/2|\phi|^4-U(x)|\phi|^2\}dx.\]
yields an effective Lagrangian
\[L_{eff}=-2BAq_t-2A(kq_t-\omega)+2(A^2+B^2)A-4/3A^3+Ak^2+2ABk\]\[\;\;\;\;+U_0A\{\tanh\{A(x_0+d-q)\}-\tanh\{A(x_0-q)\},\]
where $\omega=k^2/2+kB$ and $A^2+B^2=A_0^2=const$.
We can obtain equations of motion from the effective Lagrangian as
\begin{eqnarray}
\frac{d(BA+Ak)}{dt}&=&-A\frac{\partial U_{eff}}{\partial q},\nonumber\\
\frac{dq}{dt}&=&k+B+\frac{B}{B^2-A^2+kB}\frac{\partial (AU_{eff})}{\partial A},
\end{eqnarray}
where $U_{eff}=1/2U_0[\tanh\{A(x_0+d-q)\}-\tanh\{A(x_0-q)\}]$.  
Similar critical value of $U_0$ can be evaluated using this equation, which is almost the same as the one evaluated using the energy. However, the critical value of $U_0$ estimated by the above theory is close to half of the critical value by numerical simulations as shown in Fig.~5. The above equations of motion are approximated to $d^2q/dt^2=-\partial U_{eff}/\partial q$ when $v$ and $U_0$ are sufficiently small and then $|B|$ is small, which is the equations of motion for a classical particle of mass 1. On the other hand, Busch and Anglin derived equations of motion of the form $d^2q/dt^2=-(1/2)\partial U(q)/\partial q$  when the potential $U(x)$ changes very slowly compared to the soliton scale,\cite{Busch} using a multiple time scale boundary layer theory. In their theory, the background amplitude $A^2+B^2$ and the velocity are assumed to change slowly. The resultant effective mass of the dark soliton becomes 2. Considering the factor 1/2, it is reasonable that the numerical critical value of $U_0$ is twice of the theoretical estimate. That is, our evaluation based on the energy and the Lagrangian fails to estimate the critical value of $U_0$.  Probably, it is necessary to involve the effect of the variation of the background component $A,\,B,$ and $k$ to the Lagrangian formalism, although we have not succeeded yet.  

For a bright soliton, the kinetic energy is somewhat decreased and can be trapped in an interspace between double potential walls, even if the soliton is initially outside of the potential walls. On the other hand, the dark soliton can get some kinetic energy at the scattering and run away from the interspace between the double walls, because the dark soliton with smaller $B$-value is energetically disadvantageous. To show the detrapping, we have used the double potential walls of eq.~(4). The parameters are $U_0=0.1$, $U_1=0$, $x_1=19$ and $x_2=30$ and $d=1$. The initial amplitude $A$ is 1 and the initial velocity is $v=3\pi/50$, and the initial position is $q=25$. Figure 6(a) displays the time evolution of the dark soliton. Figure 6(b) displays the time evolution of the minimum point of $|\phi|$. Figure 6(c) displays the time evolution of the minimum value of $|\phi|$. The maximum of the $|B|$-value increases slowly each time the dark soliton is reflected by the second wall, and finally the dark soliton overcomes the potential wall. This detrapping phenomenon can be described by eq.~(9), although it takes more time to overcome the potential walls in eq.~(9). We need to improve the equations of motion to describe the detrapping phenomenon more quantitatively. 
\begin{figure}[t]
\begin{center}
\includegraphics[width=14cm]{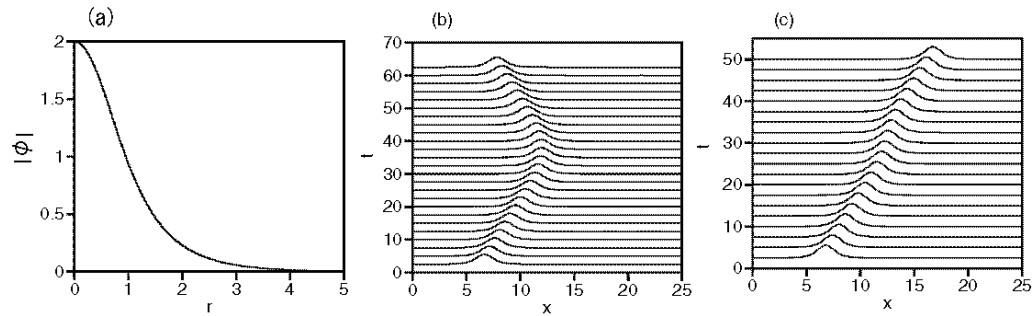}
\end{center}
\caption{(a) Amplitude profile $f(r)=|\phi(x,y,t)|$ for the two-dimensional quintic nonlinear Schr\"odinger equation with $c=0.1$.
 (b) Time evolution of $|\phi(x,L/2)|$ for the initial velocity $k_x=0.19$. (c) Time evolution of $|\phi(x,L/2)|$ for the initial velocity $k_x=0.24$.}
\label{f7}
\end{figure}
\begin{figure}[t]
\begin{center}
\includegraphics[width=9cm]{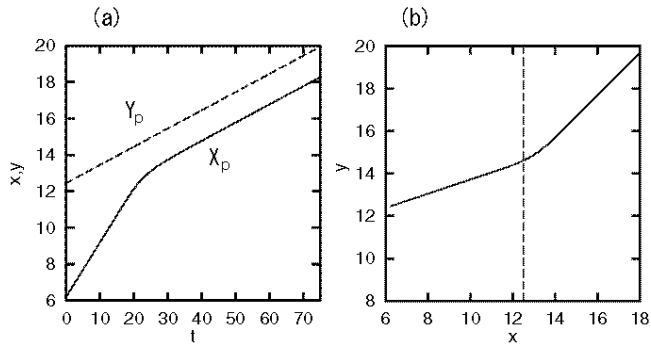}
\end{center}
\caption{(a) Time evolutions of $X_p$ and $Y_p(t)$, where $(X_p,Y_p)$ is the peak position of the two-dimensional soliton. 
 (b) Trajectory $(X_p(t),Y_p(t))$ of the peak position in the $x-y$ plane. The initial velocity is $(0.3,0.1)$. The dashed line denotes a boundary $x=L/2$ of the stepwise potential.}
\label{f8}
\end{figure}
\section{Scattering and refraction of two-dimensional solitons}
The standard two dimensional nonlinear Schr\"odinger equation does not have soliton type solutions.  Even if a localized initial condition is assumed, it may extend and changes into radiations or leads to collapse, depending on the initial conditions. However, two dimensional solitonlike solutions exist in the quintic nonlinear Schr\"odinger equation.\cite{Kiv} 
The two-dimensional quintic nonlinear Schr\"odinger equation is not an integrable system, so it may not be suitable to call the localized solution a soliton, but  we say it a two-dimensional soliton in this paper by analogy with the one-dimensional system. 
We study the scattering and the refraction of the two-dimensional soliton using the quintic nonlinear Schr\"odinger equation:
\begin{equation}
i\frac{\partial \phi}{\partial t}=-\frac{1}{2}\nabla^2\phi-|\phi|^2\phi+c|\phi|^4\phi+U(x,y)\phi.
\end{equation}
Equation (10) has an axisymmetric localized solution for $U(x,y)=0$.  
The localized solution has a form $\phi(x,y)=f(r){\rm e}^{-i\mu t}$, where $r$ is the distance from the center of the axisymmetric localized solution. The amplitude $f(r)$ needs to satisfy
\begin{equation}
f^{\prime\prime}+r^{-1}f^{\prime}=-2\{\mu+f^2-cf^4\}f.
\end{equation}
Figure 7(a) is a numerically obtained solution to the above nonlinear eigenvalue problem for $f(0)=2$ and $c=0.1$. The amplitude $f(0)$ of the two-dimensional soliton is a free parameter, as is the one-dimensional soliton. 
We have checked that the axisymmetric solution is stable by direct numerical simulations of eq.~(10). 
We can construct a propagating soliton with velocity $(v_x,v_y)=(k_x,k_y)$ using the solution of eq.~(11) as $\phi=f(x-v_xt,y-v_yt){\rm e}^{ik_x x+ik_y y-i\mu^{\prime} t}$.
We have performed numerical simulations of the quintic nonlinear Schr\"odinger equation with a potential wall of $U(x,y)=U_0=0.04$ for $L/2-0.5<x<L/2+0.5$. That is, the potential wall is parallel to the $y$-axis and has a width $d=1$.  The system size is $25\times 25$, and the initial wavenumber is $(k_x,0)$, that is, the initial velocity is perpendicular to the potential wall. Figure 7(b) and (c) display the time evolutions of $|\phi(x,L/2)|$ for $k_x=0.19$ and 0.24 at $y=L/2$. 
We have checked that the form of the two-dimensionally localized structure is maintained at the scattering. 
The two-dimensional soliton is reflected for $k_x=0.19$ and it has passed through the potential wall for $k_y=0.24$.  The critical value for the transition is $k_x=0.215$ in our numerical simulation. The Lagrangian for the quintic nonlinear Schr\"odinger equation is given by $L=\int_{-\infty}^{\infty} dxdy \{i/2(\phi_t\phi^*-\phi_t^*\phi)-1/2|\nabla\phi|^2+1/2|\phi|^4-c/3|\phi|^6-U(x,y)|\phi|^2\}$. If the solution is assumed to be $\phi(x,y)=f(x-q(t),y)e^{ip(x-q(t))-i\mu t}$, the equation of motion for the center of mass is written using the Lagrangian formalism $d/dt(\partial L_{eff}/\partial \dot{q})=\partial L_{eff}/\partial q,\,d/dt(\partial L_{eff}/\partial \dot{p})=\partial L_{eff}/\partial p$  as 
\begin{equation}
\frac{d^2q}{dt^2}=-\frac{\partial U_{eff}}{\partial q},  
\end{equation}
where $U_{eff}=\int\int f^2(x-q,y)U(x,y)dxdy/\int 2\pi rf(r)^2dr$. For the potential wall with height $U_0$ and the width $d$, the critical velocity $v_c$ is estimated from the relation
\begin{equation}
\frac{1}{2}v_c^2=\frac{\int_0^{d/2}2\pi rf^2dr+\int_{d/2}^{\infty}r4{\rm sin}^{-1}\{d/(2r)\}f^2dr}{\int_0^{\infty}2\pi r f^2dr}.
\end{equation}
For $d=1$ and $U_0=0.04$, the critical initial velocity is 0.214 for $f(r)$ shown in Fig.7(a), which is consistent with the direct numerical simulation. 
The critical energy is sufficiently smaller than the potential height $U_0$ similarly to the one-dimensional case. It might also be interpreted as the coherent tunneling. 

We have performed a numerical simulation for the steplike potential: $U(x)=0$ for $x<L/2$ and $U(x)=U_0$ for $x>L/2$. The initial velocity vector is $(k_x,k_y)=(0.3,0.1)$, and the potential height is $U_0=0.04$.  Figure 8(a) displays time evolutions of the peak position $(X_p,Y_p)$ of the two-dimensional soliton. The velocity in the $x$-direction changes from 0.3 to 0.1, and the velocity in the $y$-direction does not change. The velocity $v_x=0.1$ in the $x$-direction after the scattering satisfies the energy conservation law $1/2v_x^2+U_0=1/2\cdot 0.3^2$. As a result, the trajectory of the two-dimensional soliton is refracted, which is shown in Fig.~8(b). The angle of the refraction is nearly $\pi/4$, and it is larger than the angle of the incidence, although the velocity becomes smaller after the refraction for $x>L/2$.  This phenomenon corresponds to Newton's particle theory for the refraction. The soliton behaves like a particle, and therefore Newton's theory is revived.      
\section{Summary}
We have studied the scattering of bright solitons and dark solitons by the potential walls. We have confirmed by direct numerical simulations that the boundary line of the particlelike behavior and the wavy behavior is given by $U_{0c}=A^2/\{8\tanh(Ad/2)\}$, which was expected in our previous paper. We have found that a bright soliton can be trapped between the double potentials. When $U_0$ is small, the soliton behaves like a particle and the soliton is completely trapped in a narrow parameter region between the reflection and the transmission. When $U_0$ is large and $U_1$ is large, there appears a bound state in the interspace between the double walls.  At the collision of a soliton to the left wall, a part of the soliton is changed into the bound state and the rest is reflected.  The velocity of the reflected pulse can be faster than the initial velocity. 

We have also studied the scattering of dark solitons. When the dark soliton is reflected, the $B$-value is decreased from 0 to a certain negative value, that is, the dark soliton is changed into a grey soliton. We have evaluated the critical value of $U_0$ between the reflection and the transmission. A simple theoretical estimate is about half of the numerical values. It may be because the background density and velocity vary in space by the potential wall. The calculation by Busch and Anglin is based on the approximation of the slowly varying potential. For the potential walls whose width are comparable to or smaller than the soliton's scale, their approximation does not seem to be applicable directly, but the factor of 1/2 may be related to the effect. We need to improve the theoretical estimate by considering the effect. We have also found that a dark soliton cannot be trapped in an interspace between the double walls, and runs naturally away from the trapping region.

We have further studied the scattering and the refraction of a two-dimensional soliton in the quintic nonlinear Schr\"odinger equation.  We have found that the coherent tunneling occurs also for the two dimensional soliton. We have found a phenomenon corresponding to Newton's theory for the refraction.       

\end{document}